\pdfoutput=1

\documentclass[aps,prl,10pt,twocolumn,showpacs,superscriptaddress,footnoteinbib]{revtex4}
\usepackage{amsmath}
\usepackage{latexsym}
\usepackage{amssymb}
\usepackage{graphics,epstopdf}

\begin{document}

\title{Detecting mixedness of qutrit systems using the uncertainty relation}

\author{S. Mal}
\thanks{shiladitya@bose.res.in}
\affiliation{S. N. Bose National Centre for Basic Sciences, Salt Lake, Kolkata 700 098, India}

\author{T. Pramanik}
\thanks{tanu.pram99@bose.res.in}
\affiliation{S. N. Bose National Centre for Basic Sciences, Salt Lake, Kolkata 700 098, India}

\author{A. S. Majumdar}
\thanks{archan@bose.res.in}
\affiliation{S. N. Bose National Centre for Basic Sciences, Salt Lake, Kolkata 700 098, India}

\date{\today}

\begin{abstract}
We show that the uncertainty relation as expressed in the Robertson-Schrodinger
generalized form can be used to detect the mixedness of three-level
quantum systems in terms of measureable expectation values of suitably chosen 
observables when prior knowledge about the basis of the given state is known. 
In particular, we demonstrate the existence of observables for
which the generalized uncertainty relation is satisfied as an equality for
pure states and a strict inequality for mixed states corresponding to single
as well as bipartite sytems of  qutrits. Examples of such observables are found
for which the magnitude of uncertainty is proportional to the linear entropy
of the system, thereby providing a method for measuring mixedness.
\end{abstract}


\maketitle

\paragraph{A. Introduction.\textemdash}

The uncertainty relation lies at the heart of quantum mechanics, providing 
one of the first and foremost points of departure from classical concepts. 
As originally
formulated by Heisenberg \cite{heisen}, it prohibits certain properties of 
quantum systems from being simultaneously well-defined. A generalized form of
the uncertainty relation was proposed by Robertson \cite{Robert}
and Schrodinger \cite{Schrodinger}, and since then, several other versions
of the uncertainty principle have been suggested.   A reformulation
takes into account the inevitable noise and disturbance associated with 
measurements \cite{ozawa}. The consideration of 
state-independence has lead to  the formulation of entropic versions of
the uncertainty principle \cite{entrop}. A modification of the entropic
uncertainty relation occurs in the presence of quantum memory associated 
with quantum correlations \cite{berta}. Another version provides
a fine-grained distinction between the uncertainties inherent in obtaining
possible different outcomes of measurements \cite{oppen}.

In recent years certain important applications of uncertainty relations
have been discovered in the realm of quantum information processing.  The
security of quantum key distribution protocols is based fundamentally on
quantum uncertainty \cite{winter}, and the amount of key extractable per
state can be linked to the lower limit of entropic 
uncertainty \cite{berta,renes}.
The fine-grained uncertainty relation can be used to determine the nonlocality
of the underlying physical system \cite{oppen,pram}.
The uncertainty principle has been used for discrimination 
between separable and entangled quantum states\cite{Simon}, and 
 the Robertson-Schrodinger generalized uncertainty relation 
(GUR) has also been applied in this 
context \cite{gurent}. In the 
present work our motivation is to investigate the role of GUR in the context
of another important property, {\it viz.} the purity of quantum 
systems.

At the practical level   the ubiquitous interaction with the 
environment inevitably affects the purity of a quantum system. A relevant
issue for an experimenter is to ascertain whether a prepared pure state
has remained isolated from environmental interaction.
It becomes important to test whether a given quantum state is pure, in 
order to use it effectively as a resource for quantum information 
processing \cite{paris,pure}.
The purity of a given state is also related to the 
entanglement of a larger multipartite system of which it may be a 
part \cite{Acin}.
The mixedness of states can be characterized by the property of linear entropy,
which is  a non linear functional of the quantum state. The linear entropy
can be extracted from the given state by tomography  which 
usually is expensive in terms of resources and measurements involved.  
Bypassing  a classical evaluation process,  estimation of purity of a 
system using quantum networks has been suggested \cite{Ekert}.  Discrimination 
between pure and mixed states by positive operator valued measurements that 
amounts to a  maximum confidence discrimination, has also been 
proposed \cite{Chi}. 

In this work we connect the Robertson-Schrodinger GUR to the property
of mixedness of quantum states of discrete variables. 
For the case of continuous variable 
systems  there exist certain pure states for which the uncertainty as 
quantified by the GUR is minimized \cite{Jackiw}, and the connection of purity
with observable quantities of the relevant states have been found \cite{paris}.
Here we
show that GUR can be used to distinguish between pure and
mixed states of finite dimensional systems. To set the background we first
briefly mention the essential results 
for two-level systems. Our focus here is on three-level systems which 
are not only of
fundamental relevance in laser physics, but also the properties
of which have generated much recent interest from the perspective
of information processing \cite{distri,howell,bru,grob,klyachko,qutrit}.
We show using 
examples of single
and bipartite class of qutrit states that the GUR 
can be satisfied as an equality
for pure states while it remains an inequality for mixed states by
 the choice of suitable observables.  We prescribe an observational scheme 
using GUR which can detect mixedness of qutrit systems unambiguously,
requiring less resources compared to tomography, and  
implementable through the measurement of Hermitian witness-like operators.

\paragraph{B. GUR as a witness of mixedness.\textemdash}

GUR for any pair of observables $A,B$ and for any quantum state represented by 
the density operator $\rho$ can be written as \cite{Schrodinger,Robert}
\begin{eqnarray}
Q(A,B,\rho) \ge 0
\label{gur1}
\end{eqnarray}
where
\begin{eqnarray}
Q(A,B,\rho)&=&(\Delta A)^2 (\Delta B)^2- |\frac{\langle[A,B]\rangle}{2}|^2 \nonumber \\
&& - |(\frac{\langle\{A,B\}\rangle}{2}-\langle A\rangle \langle B\rangle)|^2
\label{gur2}
\end{eqnarray}
with $(\Delta A)^2$ and $(\Delta B)^2$ representing the variances of the 
observables, $A$ and $B$, respectively, given by
$(\Delta A)^2=(\langle A^2\rangle)-(\langle A\rangle)^2$, $(\Delta B)^2=(\langle B^2\rangle)-(\langle B\rangle)^2$, and the square (curly) brackets representing
the standard commutators (anti-commutators) of the corresponding operators.
The quantity $Q(A,B,\rho)$ involves the measurable quantities, i.e., the 
expectation values and variances of the relevant observables in the state 
$\rho$. States of a $d$-level quantum system are in one to one correspondence 
with Hermitian, positive semi-definite, unit trace operators acting on a 
$d$-dimensional Hilbert space. The defining properties of these density 
operators $\rho$ are (i) $\rho\dagger=\rho$, (ii) $\rho\geq 0  $, (iii) 
$tr[\rho]=1$. Pure states correspond to the further condition $\rho^2=\rho$ 
which is equivalent to the scalar condition $tr[\rho^2]=1 $. Hence, complement 
of the trace condition can be taken as a measure of mixedness given by
the linear entropy defined for a $d$-level system as 
\begin{eqnarray}
S_l(\rho) = (\frac{d}{d-1})(1-tr(\rho^{2}))
\label{linentrop}
\end{eqnarray}
We now investigate how the  quantity $Q(A,B,\rho)$ can act as an 
experimentally realizable measure of mixedness of a system.

We first briefly describe the status of GUR with regard to the purity of
qubit states. The density operator for two-level systems can be expressed 
in terms of the 
Pauli matrices.
The state of a single qubit can be written as
\begin{eqnarray}
\rho(\vec{n})=\frac{(I+\vec{n}.\vec{\sigma})}{2},   \>\>    \vec{n}\in \mathbb{R}^{3}
\label{singqub}
\end{eqnarray}
Positivity of this Hermitian unit trace matrix demands $ |\vec{n}|^2\leqslant1$.
It follows that single qubit states are in one to one correspondence with the 
points on or inside the closed unit ball  centred at the origin of 
$\mathbb{R}^{3}$. Points on the boundary correspond to pure states. We show that for 
a pair of 
suitably chosen spin observables,  GUR is satisfied as an equality for the 
states extremal, i.e., the pure states, and as an inequality for points other 
than extremals, i.e., for the mixed states. 
The linear entropy of the state $\rho$ can be written as
$S_l(\rho)=(1-\vec{n}^{2})$.
If we choose spin observables along two different directions, i.e., 
$A=\hat{r}.\vec{\sigma}$ and $B=\hat{t}.\vec{\sigma}$, then  $Q$ becomes
\begin{eqnarray}
Q(A,B,\rho)=(1-(\Sigma r_{i}t_{i})^{2})S_l(\rho)
\label{qub}
\end{eqnarray}
It thus follows that for $\hat{r}.\hat{t}=0$,  $Q$ coincides 
with  the linear entropy. For orthogonal spin measurements, the 
uncertainty quantified by GUR, $Q$ and the linear entropy $S_l$ are exactly 
same for single qubit systems. Thus, it turns out that $Q=0$ is both a 
necessary and sufficient condition for any single qubit system to be pure when 
the pair of observables are qubit spins along two different directions.

For the treatment of composite systems the states considered
are taken to be
polarized along a specific known direction, say, the $z$-
axis forming the Schmidt decomposition basis. The choice of $A$ and $B$, in order to enable $Q(A,B,\rho)$ as a 
 mixedness measure, 
for the two-qubit case,
are given by
\begin{eqnarray}
A=(\hat{m}.\vec{\sigma}^{1})\otimes(\hat{n}.\vec{\sigma}^{2})\nonumber\\
B=(\hat{p}.\vec{\sigma}^{1})\otimes(\hat{q}.\vec{\sigma}^{2})
\end{eqnarray}
where $\hat{m},\hat{n},\hat{p},\hat{q}$ are unit vectors.
For enabling $Z(A,B,\rho)$ to be used
for discerning the purity/mixedness of given two qubit
state specified, say, $z$-axis, the appropriate choice of observables 
$A$ and $B$ is found to be that of lying on the two
dimensional $x-y$ plane (i.e.,$\hat{m},\hat{n},\hat{p},\hat{q}$ are all taken
to be on the $x-y$ plane), normal to the $z$-axis pertaining
to the relevant Schmidt decomposition basis.  Then, $Q(A,B,\rho)=0$ 
(i.e., GUR is satisfied as an equality) necessarily holds good for
pure two-qubit states whose individual spin orientations
are all along a given direction (say, the $z$-axis) normal to
which lies the plane on which the observables $A$ and $B$
are defined. On the other hand, $Q(A,B,\rho) > 0$ holds good for most settings 
of $A$ and
$B$ for two qubit isotropic states, for the Werner class of states given by
$\rho_{w}=((1-p)/4)I+p\rho_{s}$
 ($\rho_s$ is the two-qubit singlet state),
as well for other types of one parameter two-qubit states which comprise of 
pure states whose
individual spin orientations are all along the same given
direction normal to the plane on which the observables
$A$ and $B$ are defined.
 For the case of multipartite systems, in our purpose the general form of $n$-qubit
observables is given by 
\begin{eqnarray}
A=\hat{r}_1.\vec{\sigma}\otimes\hat{r}_2.\vec{\sigma}\otimes...\otimes\hat{r}_n.\vec{\sigma}\nonumber \\
B=\hat{t}_1.\vec{\sigma}\otimes\hat{t}_2.\vec{\sigma}\otimes...\otimes\hat{t}_n.\vec{\sigma}
\end{eqnarray}
where, $\hat{r}_i,\hat{t}_i$ are unit vectors in $\mathbb{R}^{3}$.  GUR may be 
used to distinguish pure states from mixed ones  with the choice of suitable 
observables for composite
qubit systems, whose detailed implications will be presented in a
separate work.

\paragraph{C. Three-level systems.\textemdash}

The structure of the state space of the generalised Bloch sphere ($\Omega_{d}$),
is much richer for $d\geq3$ \cite{mukunda,Goyal}. 
Qutrit states can be expressed in terms of 
Gellmann matrices that are familiar generators of the unimodular 
unitary group $SU(3)$ in its defining representation with
eight Hermitian, traceless and orthogonal 
matrices $\lambda_j, j=1,....,8$ satisfying
$tr(\lambda_{k}\lambda_{l})=2 \delta_{kl}$, and
$\lambda_{j}\lambda_{k}=(2/3) \delta_{jk}+ d_{jkl} \lambda_{l}+i f_{jkl} \lambda_{l}$.
The expansion coefficients $f_{jkl}$, the structure constants of the Lie algebra
 of $SU(3)$, are totally anti-symmetric, while $d_{jkl}$ are totally symmetric. 
Single-qutrit
states can be expressed as
\begin{eqnarray}
\rho(\vec{n})=\frac{I+\sqrt{3}\vec{n}.\vec{\lambda}}{3}, \vec{n}\in \mathbb{R}^{8.}
\end{eqnarray}
The set of all extremals (pure states) of $ \Omega_{3} $ constitute also 
$CP^{2}$, and can be written as 
$\Omega_{3}^{ext}=CP^{2}=\{\vec{n}\in \mathbb{R}^{8}|\vec{n}.\vec{n}=1, \vec{n}*\vec{n}=\vec{n}\}$, with 
$\vec{n}*\vec{n}=\sqrt{3}d_{jkl}n_{k}n_{l}\hat{e_{j}}$.
Here $\hat{e_{j}}$ is the unit vector belongs to $\mathbb{R}^{8}$.
Non-negativity of $\rho$ demands that $\vec{n}$ should satisfy the additional 
inequality $|\vec{n}|^2\leqslant1$. The
boundary $\partial\Omega_{3}$ of $\Omega_{3}$ is characterised by
$\partial\Omega_{3}=\{\vec{n}\in \mathbb{R}^{8}| 3\vec{n}.\vec{n}-2\vec{n}*\vec{n}.\vec{n}=1,\vec{n}.\vec{n}\leqslant 1\}$, and 
the state space $\Omega_{3}$ is given by 
$\Omega_{3}=\{\vec{n}\in \mathbb{R}^{8}|3\vec{n}.\vec{n}-2\vec{n}*\vec{n}.\vec{n}\leqslant 1, \vec{n}.\vec{n}\leqslant 1\}$.
For two-level systems the whole boundary of the state space represents pure 
states, i.e., $\Omega_{2}^{ext}=\partial\Omega_{2}$, while for three-level systems
 $\Omega_{3}^{ext}\subset\partial\Omega_{3}$. The four parameter family 
$\Omega_{3}^{ext}$ is sprinkled over the seven parameter surface 
$\partial\Omega_{3}$ of $\Omega_{3}$.

 The most general type of 
observables can be written as ${A}=\hat{a}.\vec{\lambda}=a_{i}\lambda_{i}$, ${B}=\hat{b}.\vec{\lambda}=b_{i}\lambda_{i}$, where, $\Sigma a_{i}^{2}=1$ and $\Sigma b_{i}^{2}=1$. 
The measurement of 
qutrit observables composed of the various $\lambda_i$'s,  can be recast 
in terms of qutrit spin observables, given by \cite{klyachko}, e.g.,
$\lambda_{1} = (1/\sqrt{2})(S_{x}+2 \{S_{z},S_{x}\})$, and similarly
for the other $\lambda_i$'s. 
Where the qutrit spins are given by 
\begin{eqnarray}
\sqrt{2}S_{x}=\left(\begin{array}{ccc}
0 & 1 & 0 \\ 
1 & 0 & 1 \\ 
0 & 1 & 0
\end{array}\right), \sqrt{2}S_{y}=\left(\begin{array}{ccc}
0 & -i & 0 \\ 
i & 0 & -i \\ 
0 & i & 0
\end{array}\right),\nonumber\\ S_{z}=\left(\begin{array}{ccc}
1 & 0 & 0 \\ 
0 & 0 & 0 \\ 
0 & 0 & -1
\end{array}\right). 
\end{eqnarray}\\
Note that with the choice of $A=\hat{A}.\hat{\lambda}$ and 
$B=\hat{B}.\hat{\lambda}$, $Q$ becomes
\begin{eqnarray}
Q&=&(4/9)(1-(\hat{A}.\hat{B})^2)+(4/9)(((\hat{A}\ast\hat{A}).\vec{n})+((\hat{B}\ast\hat{B}).\vec{n})\nonumber\\
&-&2(\hat{A}.\hat{B})((\hat{A}\ast\hat{B}).\vec{n}))
+(4/9)(((\hat{A}\ast\hat{A}).\vec{n}) ((\hat{B}\ast\hat{B}).\vec{n})\nonumber\\
&-&((\hat{A}\ast\hat{B}).\vec{n})^2
+4(\hat{A}.\hat{B})(\hat{A}.\vec{n})( \hat{B}.\vec{n})-2(\hat{A}.\vec{n})^2-2(\hat{B}.\vec{n})^2\nonumber\\
-&3&((\hat{A}\wedge\hat{B}).\vec{n})^2)
-(4/9)(2((\hat{A}\ast\hat{A}).\vec{n})) (\hat{B}.\vec{n})^2\nonumber\\
&+&2(\hat{A}.\vec{n})^2 ((\hat{B}\ast\hat{B}).\vec{n}))-4((\hat{A}\ast\hat{B}).\vec{n})(\hat{A}.\vec{n})( \hat{B}.\vec{n}))
\end{eqnarray}
where $(\hat{A}\ast\hat{B})_{k}=\sqrt{3}d_{ijk}A_{i}B_{j}$ and $(\hat{A}\wedge\hat{B})_{k}=f_{ijk}A_{i}B_{j}$. From the expression of $Q$ it is clear that it changes
 if $\rho$ is changed by some unitary transformation.  For such change of 
states the norm of $\vec{n}$ does not change. Purity/mixedness property of a 
state does not change under unitary operations on the state. Hence, it is 
desirable for any mixedness measure to remain invariant under unitary 
operation. This would be possible if $Q$ becomes some function of only 
$|\vec{n}|^2$ for suitable choice of observables. However, unlike the case of 
the single qubit, for the single qutrit $Q$ becomes independent of the
linear and cubic terms of $|\vec{n}|$ only for the trivial 
choice of observables, i.e., $\hat{A}=\hat{B}$, in which case $Q$ becomes
zero, whatever be the state, pure or 
mixed. Here we employ  suitably chosen observables and a sequence of 
measurements to turn $Q$ to a detector of mixedness, i.e., $Q=0$ for pure,
and $Q>0$ for mixed states. Note further, that 
under a basis transformation $\lambda_{i}'=U\lambda_{i}U^{\dagger}$, the
state becomes $\rho'=(1/3)(I+\sqrt{3}\vec{n}'.\vec{\lambda}')=U(1/3)(I+\sqrt{3}\vec{n}'.\vec{\lambda})U^{\dagger}$. Now, for any observable $\chi'$ in the prime basis,
one has $Tr[\chi'\rho']= Tr[\chi (1/3)(I+\sqrt{3}\vec{n}'.\vec{\lambda})]$. Thus,
any non-vanishing expectation value in the primed basis cannot vanish in
the unprimed one, and vice-versa.  Hence, in order to measure in another 
basis one has to simply choose observables which are unitary conjugates to the 
observables written in terms of standard $\lambda$ basis. Such observables
would again yield $Q=0$ for pure,
and $Q>0$ for mixed states in the new basis. Hence, though
we have specified our scheme based on the single qutrit state in terms of 
the standard $\lambda$
basis \cite{mukunda,Goyal}, our scheme remains invariant with regard to the
choice of the basis as long as the knowledge of the specific basis chosen
is available to the experimenter. This means that the experiment shall involve not only the observables $A$
and $B$ but also a possibility for simultaneous unitary rotations of these observables.

In what follows we take up to three-parameter 
family of states from 
$\Omega_{3}$ \cite{Goyal}, and find that there exist 
observable pairs which for pure  states exhibit minimum uncertainty,
{\it viz.} $Q=0$.  Our scheme runs as follows.
 Economizing on the number of 
measurements required, we take  $\lambda_{3}$ as $A$ and sequentially, the 
members of any one of the pairs $(\lambda_{7},\lambda_{6}),(\lambda_{5},\lambda_{4}),(\lambda_{1},\lambda_{2})$ as $B$. The significance of such pairing will be 
clear later. If two successive measurements taking $B$ from any of the above
 pairs yield $Q=0$, the state concerned is pure. In contrast, if $B$ taken from
all the above pairs sequentially, yields $Q>0$, the state is found to be mixed.
(See, Fig.1 for an illustration of the scheme).

\begin{figure}[!ht]
\resizebox{8cm}{7cm}{\includegraphics{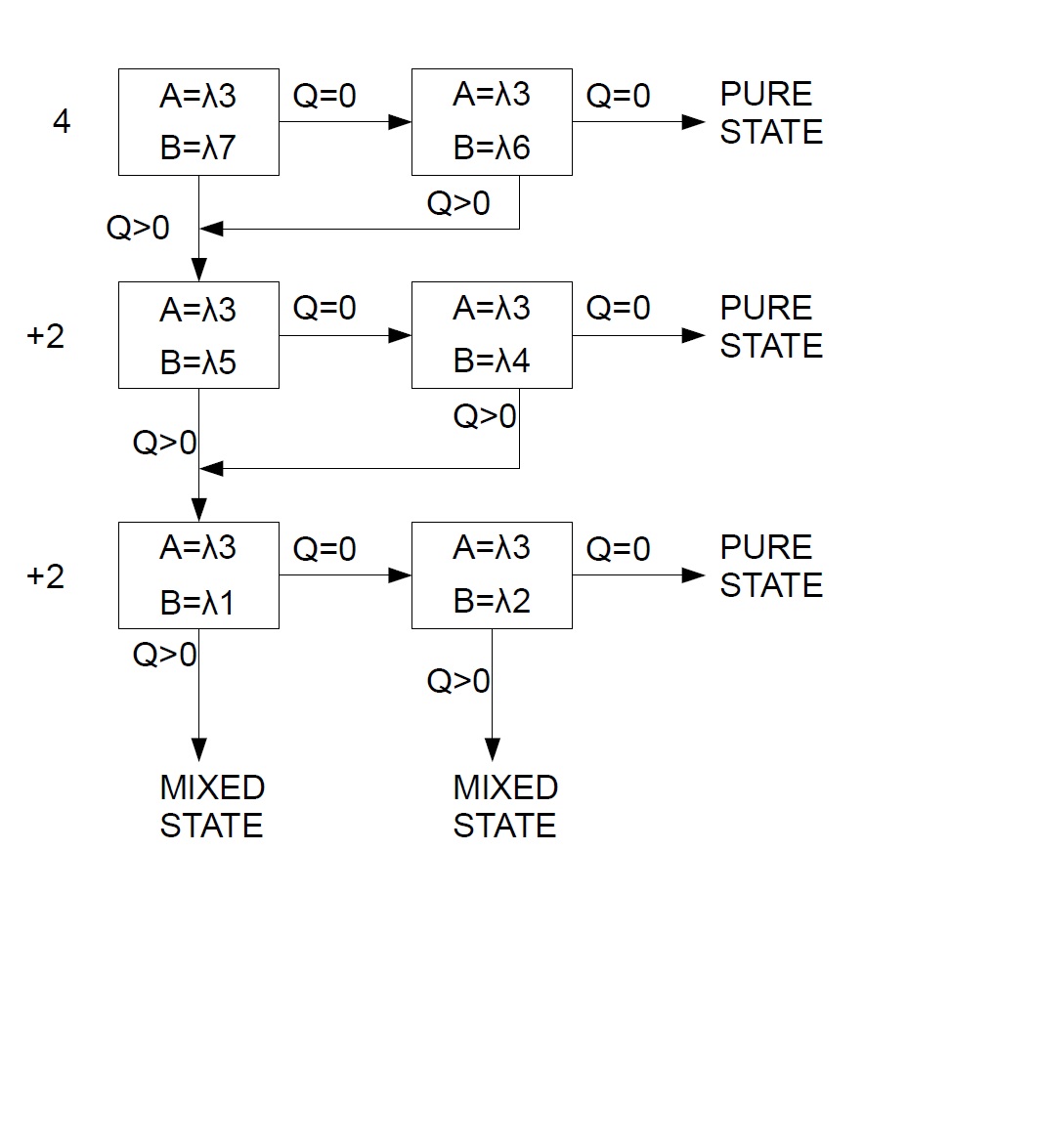}}
\caption{\footnotesize Detection scheme for purity of single qutrit states of up to 
three parameters. The numers to the left of the boxes indicate the number
of measurements required corresponding to each of the horizontal levels.}
\end{figure}

Let us first consider the one-parameter family of single-qutrit states
 for which only one of the eight parameters 
($n_{i}, i=1,...,8$) is non-zero while the remaining seven vanish.  
The linear entropy of this class of states is given by
\begin{equation}
S_l(\rho)=1-n_{i}^{2}
\label{linsing}
\end{equation}
There exist many pairs of observables which can 
detect mixedness of this class of states unambiguously.
For example, when $i=8$, the only pure state of this class is
given by $n_{8}=-1$ \cite{Goyal}. Here
\begin{equation}
Q(\lambda_{3},\lambda_{7})=Q(\lambda_{3},\lambda_{6})=(4/9)(2-n_{8})(1+n_{8})
\end{equation}
Hence, $Q=0$ only for  $n_{8}=-1$, but $Q>0$ otherwise. Next,
for example when $i=1$, one has
\begin{equation}
Q(\lambda_{3},\lambda_{7})=Q(\lambda_{3},\lambda_{6})=Q(\lambda_{3},\lambda_{5})=Q(\lambda_{3},\lambda_{4})=4/9
\end{equation}
It turns out that there is no choice of $B$ from both the sequential pairs
(as depicted in Fig.1)
for which $Q=0$. Similar considerations are valid also for other
single parameter qutrit states, enabling the detection scheme as given in Fig.1.
 
Moving to the two-parameter family of density matrices, (two of the eight 
parameters $n_{1}...n_{8}$ are non zero, while remaining six vanish), note that 
in this case there are twenty-eight combinations of different pairs of non-zero
 parameters, and these classes belongs to one of the four different types of 
unitary equivalence classes, {\it viz.}, circular, parabolic, elliptical and 
triangular \cite{Goyal}. In this case, for example, for states belonging to 
the parabolic class, by choosing $n_{3}$ and $n_{4}$ to be non-vanishing, $Q$ 
takes the forms
\begin{eqnarray}
Q(\lambda_{3},\lambda_{5})&=&(2/9)(2+\sqrt{3}n_{3})(1-2n_{3}^{2})-n_{4}^{2}/3 \nonumber\\
 Q(\lambda_{3},\lambda_{4})&=&(1/9)(4-8n_{3}^{2}-4\sqrt{3}n_{3}^{3}-11n_{4}^{2}\nonumber\\
&+&2\sqrt{3}n_{3}(1+4n_{4}^{2})) 
\end{eqnarray}
 Here pure states occur for  $(n_{3},n_{4})=(1/\sqrt{3},\pm\sqrt{2/3})$, leading
to $Q=0$, while $Q>0$ corresponding to all mixed states, as is also evident
from the expression for the linear entropy given by
\begin{equation}
S_l(\rho)=(1-n_{3}^{2}-n_{4}^{2})
\end{equation}
Similar considerations apply to other single qutrit states of the two parameter
family, enabling the detection of pure states  when two successive 
measurements with $B$ taken from sequential pairs (Fig.1) lead to $Q=0$.

Next consider the three-parameter family of qutrit states where there are 
seven geometrically distinct and ten unitary equivalent types of three-sections
 out of fifty-six standard three-sections. Considering an example of states 
belonging to the parabolic geometric shape, $Q$ has the forms 
\begin{eqnarray}
Q(\lambda_{3},\lambda_{5})&=&(1/9)(4-8n_{3}^{2}-4\sqrt{3}n_{3}^{3}-3n_{4}^{2}-11n_{5}^{2}\nonumber\\
&+&2\sqrt{3}n_{3}(1+4n_{5}^{2}))\nonumber\\
Q(\lambda_{3},\lambda_{4})&=&(1/9)(4-8n_{3}^{2}-4\sqrt{3}n_{3}^{3}-3n_{5}^{2}-11n_{4}^{2}\nonumber\\
&+&2\sqrt{3}n_{3}(1+4n_{4}^{2})) 
\end{eqnarray}
The linear entropy of this class of states is given by
\begin{equation}
S_l(\rho)=1-n_{3}^{2}-n_{4}^{2}-n_{5}^{2}
\label{linthree}
\end{equation} 
When  $B$ is chosen from the $(\lambda_{4},\lambda_{5})$  pair as above,  $Q$ 
turns out to be zero for pure states given by $n_{3}=1/\sqrt{3}$ and $n_{4}^{2}+n_{5}^{2}=2/3$, and  $Q$ is greater than zero for all mixed states. It can be
checked that the purity of all three parameter family of single qutrit states
can be determeined by the scheme depicted in Fig.1. 

Let us now discuss the case of two-qutrit state discrimination. Here we assume
that the states considered are taken to be
polarized along a specific known direction, say, the $z$-
axis forming the Schmidt decomposition basis.
A two-qutrit pure state in the Schmidt form can be written as 
$\vert\psi\rangle= k_{1}\vert 11\rangle+k_{2}\vert 22\rangle+k_{3}\vert 33\rangle$
where, $k_{1},k_{2},k_{3}$ are real with $k_{1}^{2}+k_{2}^{2}+k_{3}^{2}=1$, and
$\vert1\rangle$, $\vert2\rangle$ and $\vert3\rangle$ 
are orthonormal unit vectors in $\mathbb{C}^{3}$.
In our purpose a general form of observables acting on the two-qutrit system is given 
by $A=\hat{r}_1.\vec{\lambda}\otimes\hat{r}_2.\vec{\lambda}$, and
$B=\hat{t}_1.\vec{\lambda}\otimes\hat{t}_2.\vec{\lambda}$,
where $\hat{r}_1,\hat{t}_1,\hat{r}_2,\hat{t}_2$ are unit vectors in $\mathbb{R}^{8}$.
For our purpose it is sufficient to take observables of the form
\begin{eqnarray}
A &=& \lambda_{i}\otimes(\cos\theta_{2}\lambda_{i}+\sin\theta_{2}\lambda_{j})\nonumber\\
B &=& (\cos\theta_{3}\lambda_{i}+\sin\theta_{3}\lambda_{j})\otimes(\cos\theta_{4}\lambda_{i}+\sin\theta_{4}\lambda_{j})
\label{obserqut2}
\end{eqnarray}
  where $(i,j)$ are taken from the pair $(1,2)$,$(3,8),(4,5),(6,7)$, and 
$\theta_{2}, \theta_{3},\theta_{4}$ are angles between  $\hat{r}_1$ and 
$\hat{r}_2, \hat{t}_1, \hat{t}_2$, respectively.
  With the choice of observables ($i=1, j=2$), the uncertainty becomes 
  $Q(A,B,\rho_{pure})=4 k_{1}^{2} k_{2}^{2} k_{3}^{2} \sin({\theta_{2}-\theta_{3}-\theta_{4}})$.
  Hence, choosing $\theta_{2}-\theta_{3}=\theta_{4}$, we can make $Q=0$ 
for every pure state.
 
Now consider a one-parameter class of two-qutrit mixed states 
expressed as 
\begin{eqnarray}
\rho_{m}=p\rho_{1}+(1-p)\rho_{2}
\label{twoqutgen}
\end{eqnarray} 
where $\rho_{1}$ and $\rho_{2}$ are 
arbitrary pure states parametrized as $\rho_1 = |\psi_1\rangle\langle\psi_1|$
with $\vert\psi_1\rangle= k_{1}\vert 11\rangle+k_{2}\vert 22\rangle+k_{3}\vert 33\rangle$, and $\rho_2 = |\psi_2\rangle\langle\psi_2|$ with $\vert\psi_2\rangle= k_{4}\vert 11\rangle+k_{5}\vert 22\rangle+k_{6}\vert 33\rangle$. For such states the linear entropy is given by 
\begin{eqnarray}
S_l(\rho_{m})=\frac{3}{2}p(1-p)
\end{eqnarray}
The expression for $Q$ under the condition $\theta_{2}-\theta_{3}=\theta_{4}$ is 
given by
$Q(A,B,\rho_{m}) = 4 k_{1}^{2} p(1-p)(1-k_6^2 - 4k_4^2k_5^2(1-p) \cos^2(\theta_{3}+\theta_{4}))\sin^{2}(\theta_{3})$ which when
maximized over all observables in the selected region ($i=1,j=2$) leads to
\begin{eqnarray}
Q=4 k_{1}^{2}(1-k_{6}^{2})p(1-p)
\end{eqnarray}
We observe that the expression for the uncertainty may coincide with the value 
of linear entropy for certain choices of the state parameters. In general, $Q$
always vanishes for pure states, and remains positive for mixed ones, for 
$k_{1} \neq 0$, and $k_6 \neq 1$.

As another example of two-qutrit states, we consider the popular class of
isotropic states that are invariant under the action of local unitary 
operations of the form $U\otimes U^{*}$. Two-qutrit isotropic states can be 
written as
\begin{eqnarray}
\rho=p\rho_{i}+ \frac{1-p}{9}I\otimes I
\end{eqnarray}
where, $0\le p \le 1$, and  $\rho_{i}=\vert\phi\rangle\langle\phi\vert $,
with
$|\phi_{i}\rangle=(1/\sqrt{3})(|11\rangle +|22\rangle + |33\rangle)$.
The linear entropy of this state is given by 
\begin{eqnarray}
 S_l(\rho)=\frac{2}{3}(1-p^{2})
\label{liniso}
\end{eqnarray}
and our choice of observables  leads to
$Q= (8/81)(-1+p)(-3-3 p+2 p^{2}+(-1+p)\cos(2\theta_{3})
 +2 p^{2}\cos(2(\theta_{3}+\theta_{4})))^{2}\sin\theta_{3}$.
Maximizing over all observables in the selected region we get 
\begin{eqnarray}
Q=\frac{16}{81}(1-p)(1+2 p)
\end{eqnarray}
which is quadratic in the parameter $p$ similar to the linear entropy, and is
able to distinguish mixed states from the pure state ($p=1$). It may be noted
that for the Werner class of states that are invariant under the local unitary 
operations of the form $U\otimes U$, and which differ from the Isotropic class
for qutrits, there exists no pure state for qutrits, a fact that is reflected
in the corresponding expression for $Q$ that turns out to be $Q>0$ always.

\paragraph{D. Measurement prescription.\textemdash}

We now outline our suggested scheme for using the uncertainty 
relation to determine whether a given state is pure or mixed, provided
the prior knowledge of the basis is available. The generalized uncertainty
relation through the scheme discussed here is able to distinguish between
pure and mixed states for a broad category of two- and three-level systems
(see Fig.2). For single party systems, the scheme works for all qubits
and up to three-parameter family of qutrit states for which the classification
into unitary equivalence classes is available in the literature 
\cite{mukunda,Goyal}. For bipartite systems, the scheme has been shown to work
for the mixture of two arbitrary pure states, the isotropic class, and the
Werner class of states, as well. There may of course exist other classes of
states within the above categories, for which we are yet to ascertain the 
viability of this scheme.

\begin{figure}[!ht]
\resizebox{9cm}{11cm}{\includegraphics{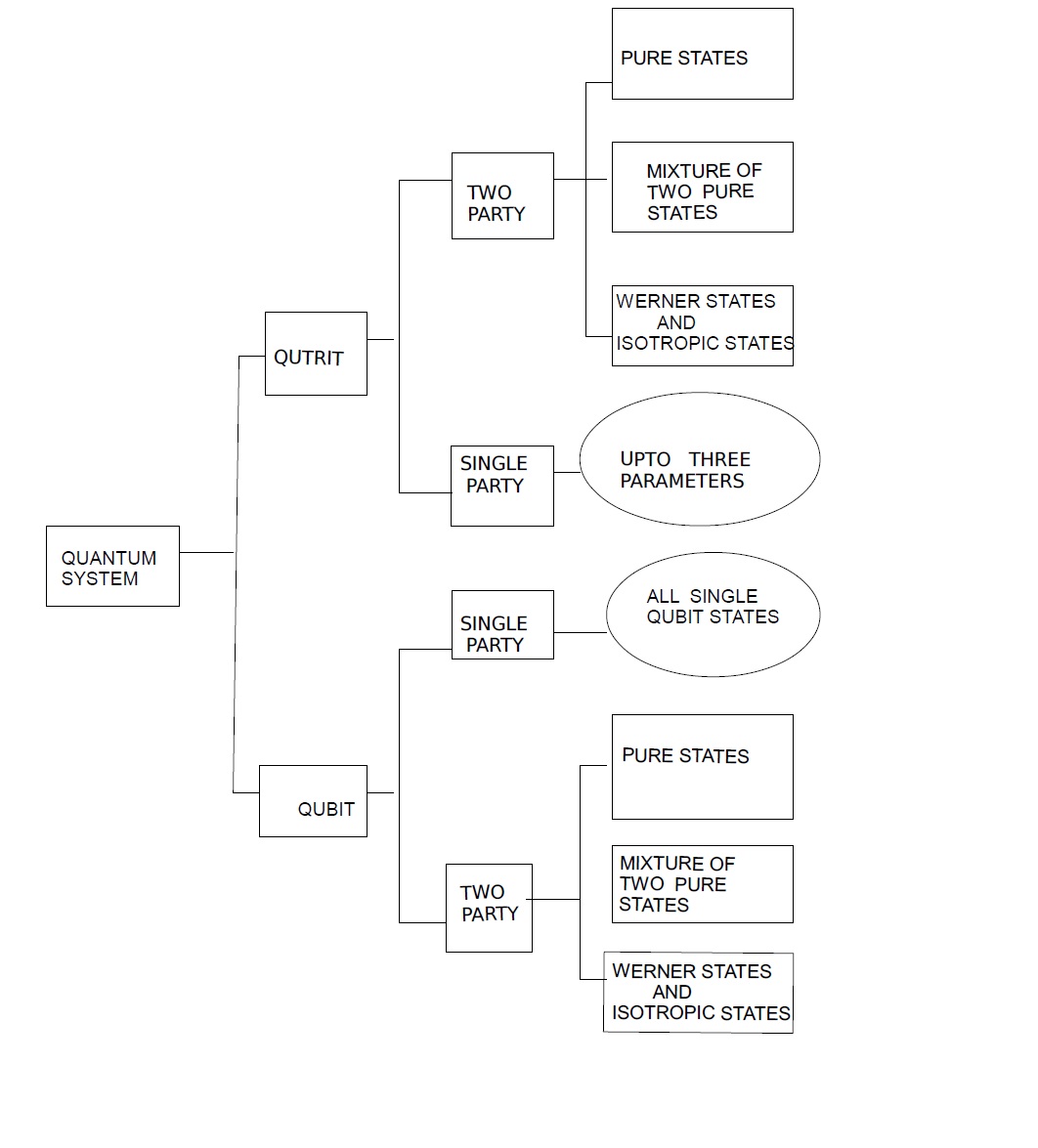}}
\caption{\footnotesize Family of states that can be distinguished using the uncertainty relation.}
\end{figure}

It may be noted here that the limitation of instrumental precision 
could  make the 
observed value of $Q$ for
pure states to be a small number in stead of exactly zero. In order to
take into account the experimental inaccuracy, a parameter  
$\varepsilon$ may be introduced in the analysis. For a single-qubit system, 
by choosing the measurement
settings for  $A$ and $B$ as qubit spins along $z$ and $x$ 
directions, respectively,   the measured value of the uncertainty  
obtained as $Q\ge\varepsilon$ leads to the conclusion that  
the given state is mixed. This prescription
of determining mixedness holds for all single-qubit states 
$\rho(\vec{n})=\frac{(I+\vec{n}.\vec{\sigma})}{2}$, except those lying
in the narrow range $1 \ge n \ge \sqrt{1- 2 \varepsilon/3}$, as determined by 
putting $Q < \varepsilon$ in Eq.(\ref{qub}).

A somewhat more elaborate procedure is required for qutrits, as may be expected 
from the richer structure of their state space. For the case of single
qutrits belonging to the one, two or three-parameter family of states, one has 
to find $Q$ taking $A=\lambda_{3}$ and $B$ from the $(\lambda_{6},\lambda_{7}),(\lambda_{4},\lambda_{5}),(\lambda_{1},\lambda_{2})$ pairs in succession as depicted 
in the Fig. 1. If $Q < \varepsilon$  for the settings $B$ corresponding to both 
members of a same pair measured in succession, then the state is pure within the limitations of experimental accuracy.  Whenever $Q\ge \varepsilon$, $B$ is
chosen from the pair vertically below. If there 
exists no such pair for which  $Q < \varepsilon$ , then the state is mixed.  In order to 
maximize
the uncertainty measured by the variable $Q$, such that $Q \ge \varepsilon$ 
for the maximum number of mixed states, the observables need to chosen so
as to avoid $|a_i/b_i| \approx 1$. Our scheme pictorially represented 
in Figure 1  is able to detect mixedness of single qutrit states up to
three parameters.

For the case of two-qutrit states,
the measurement of the observables given by Eq.(\ref{obserqut2}) with the 
$\lambda_i$'s chosen
from  the regions spanned by ($\lambda_1,\lambda_2$), together with the
restriction on the angle $\theta_3 \neq \pi$,
 suffices to distinguish pure and mixed states.
Such a procedure is able to detect all mixed states within the margin of
experimental accuracy. For example, for the case of the two-qutrit isotropic
states the method would fail only for 
states lying in
the parameter range $\sqrt{1-3\varepsilon/2} < p < 1$.

The determination of mixedness using GUR may
require in certain cases a considerably lesser number of measurements compared 
to tomography.
In the case of single  
qutrit states, full tomography involves the estimation of eight parameters,
while in our prescription sometimes four  measurements may suffice for
 detecting purity of a single qutrit state.
In figure 1, the numbers besides boxes indicate the numbers of measurements 
required to find the various expectation values including those of
(anti-)commutators required to determine $Q$ using Eq.({\ref{gur2}). For 
instance, the number $4$ besides the top box, means 
that the four measurements ($\langle\lambda_{3}\rangle,\langle\lambda_{7}\rangle$,$\langle\lambda_{8}\rangle$ and$\langle\lambda_{6}\rangle$) are all that
is required for the first horizontal level. This follows from the algebra
\begin{eqnarray}
\langle\lbrace\lambda_{3},\lambda_{7}\rbrace\rangle=-\langle\lambda_{7}\rangle/2, \nonumber\\
\langle[\lambda_{3},\lambda_{7}]\rangle=\langle\lambda_{6}\rangle/2,\nonumber\\
 \langle\lambda_{3}^{2}\rangle=\frac{2}{3}I+\frac{1}{\sqrt{3}}\langle\lambda_{8}\rangle, \nonumber\\
\langle\lambda_{7}^{2}\rangle=\frac{2}{3}I-\frac{1}{2\sqrt{3}}\langle\lambda_{8}\rangle-\frac{1}{2}\langle\lambda_{3}\rangle, \nonumber\\
\langle[\lambda_{3},\lambda_{6}]\rangle=-\langle\lambda_{7}\rangle/2,\nonumber\\
\langle\lbrace\lambda_{3},\lambda_{6}\rbrace\rangle=-\langle\lambda_{6}\rangle/2\nonumber\\
\langle\lambda_{6}^{2}\rangle=\frac{2}{3}I-\frac{1}{2\sqrt{3}})\langle\lambda_{8}\rangle-\frac{1}{2}\langle\lambda_{3}\rangle
\end{eqnarray}
To proceed vertically down to the next level in Fig.1, the number of extra 
measurements are indicated besides the boxes. It may  be mentioned that in 
our scheme it does not matter if any horizontal pair of boxes are interchanged 
with another pair at a different level. A maximum of eight measurements thus
suffices to distinguish between pure and mixed states of single qutrit up 
to three-parameter families. The maximum number of measurements required
in particular cases may not provide a significant advantage over tomography,
but would still form an independent check of states with prior knowledge
of basis. The difference in the number of required measurements is substantially
enhanced for composite states. For two qubits, GUR requires up to 
five measurements
compared to fifteen required by tomography for the class of states considered. 
For the case of two-qutrits 
the measurement of at most eight expectation
values, {\it viz.}, $\langle \lambda_1 \otimes \lambda_1 \rangle $,  $\langle \lambda_1 \otimes \lambda_2 \rangle $, $\langle \lambda_2 \otimes \lambda_1 \rangle $, $\langle \lambda_2 \otimes \lambda_2 \rangle $, $\langle \lambda_3 \otimes \lambda_3 \rangle $, $\langle \lambda_3 \otimes \lambda_8 \rangle $, $\langle \lambda_8 \otimes \lambda_3 \rangle $, and $\langle \lambda_8 \otimes \lambda_8 \rangle $, suffices using GUR for the observables defined by Eq.(\ref{obserqut2}).
A comparison of the number of measurements required using GUR with that needed
in tomography is provided in Table 1.

\begin{table}
\begin{tabular}{|c|c|c|}
\hline  System &  in tomography &  using GUR\\
\hline  Single qubit  & 3 & 3 \\
\hline  two qubit&15  & 3-5  \\
\hline  Single qutrit&8  &4-8  \\
\hline Two Qutrit & 80 & 4-8 \\
\hline
\end{tabular}\\
\caption{A comparison between the number of measurements required in tomographic method and in our method is shown for the categories of states considered. Number of measurements for detecting mixedness/purity for bipartite system is much less and for single party system this method is becoming advantageous with increasing dimension.}
\end{table}

\paragraph{E. Conclusions.\textemdash}

We have shown that the
Robertson-Schrodinger uncertainty relation \cite{Robert,Schrodinger} 
is connected to the property of mixedness
of  single and bipartite  three-level quantum systems. The
generalized uncertainty corresponding to the measurement of suitable observables
vanishes for pure states  and is positive definite for mixed states. Using this
feature we have proposed a scheme to distinguish pure and mixed states 
belonging to the classes of  all single-qutrit states up
to three parameters, as well as several classes of
two-qutrit states, when prior knowledge of the basis is available. Since the 
class of all pure states is not convex, 
the witnesses proposed here for detecting mixedness do not arise from the 
separability criterion that holds for the widely studied entanglement witnesses
\cite{guhne} as well as the recently proposed teleportation witnesses
\cite{ganguly}.   Nonetheless, the same principle of distinction of categories
of quantum states based on the measurement of expectation values of hermitian
operators is followed. 

A possible implementation of the witnesses proposed
here could be through techniques involving measurement of two-photon 
polarization-entangled modes for qutrits \cite{howell,grob}. The procedure 
suggested here could be helpful also for the detection of entanglement, since
purity of subsystems is related to the entanglement of the joint system. 
The method of detecting mixedness using the uncertainty relation is 
advantageous over tomography in terms of the number of measurements required,
siginificantly for bipartite qutrit systems, which may have applications in 
information
processing protocols such as distributed computing \cite{distri} and
security enhancement of quantum cryptography \cite{bru,grob}.

{\it Acknowledgements:}  We would like to thank Dipankar Home and Peter Holland 
for discussions and suggestions which lead to this
work.  ASM
and SM acknowledge support from the DST project
SR/S2/PU-16/2007.

\end{document}